\documentclass[a4paper,fleqn,usenatbib]{mnras}
\usepackage[T1]{fontenc}
\usepackage{ae,aecompl}
\usepackage{graphicx}	
\usepackage{amsmath}	
\usepackage{amssymb}	

\title[Demography of galactic technosignatures]{Demography of galactic technosignatures}

\author[C. Grimaldi]{
Claudio Grimaldi,$^{1}$\thanks{E-mail: claudio.grimaldi@epfl.ch}
\\
$^{1}$Laboratory of Physics of Complex Matter, Ecole Polytechnique F\'ed\'erale
de Lausanne, Station 3, CP-1015 Lausanne, Switzerland}

\pubyear{2020}

\begin{document}
\label{firstpage}
\pagerange{\pageref{firstpage}--\pageref{lastpage}}
\maketitle

\begin{abstract}
Probabilistic arguments about the existence of technological life beyond Earth traditionally refer to the Drake equation to draw possible 
estimates of the number of technologically advanced civilizations releasing, either intentionally or not, electromagnetic emissions in the
Milky Way. Here, we introduce other indicators than Drake's number $N_D$ to develop a demography of artificial 
emissions populating the Galaxy. We focus on three main categories of statistically independent signals (isotropic, narrow beams, and rotating beacons) to calculate
the average number $N_G$ of emission processes present in the Galaxy and the average number of them crossing Earth, $\bar{k}$,
which is a quantity amenable to statistical estimation from direct observations.
We show that $\bar{k}$ coincides with $N_D$ only for isotropic emissions, while $\bar{k}$ can be orders of magnitude smaller than $N_D$
in the case of highly directional signals.
We further show that while $N_D$ gives the number of emissions being released at the present time, $N_G$ considers also the signals from no
longer active emitters but whose emissions still occupy the Galaxy. We find that as long as the average longevity of the emissions is shorter
than about $10^5$ yr, $N_G$ is fully determined by the rate of emissions alone, in contrast to $N_D$ and $\bar{k}$ which depend also
on the emission longevity. Finally, using analytic formulas of $N_G$, $N_D$, and $\bar{k}$ determined for each type of emission processes 
here considered, we provide a comprehensive overview of the values these quantities can possibly achieve as functions of the emission 
birthrates, longevities, and directionality. 
\end{abstract}




\section{Introduction}
\label{intro}

As new extrasolar worlds are being routinely discovered, there is an ever mounting evidence that a significant fraction of exoplanets may have
environmental conditions suitable for developing life \citep{Dressing2013,Petigura2013,Zink2019}. In the hunt for signs of life beyond the solar 
system, the search for biosignatures from the
atmosphere and surface of extrasolar Earth-like planets is moving its first steps and will likely dominate the exoplanet science in the next  
decades \citep{Kiang2018}. The prospects of life elsewhere in the Galaxy have also reinvigorated the longstanding search for putative signs of technologically
 savvy life from beyond Earth. The search for such technosignatures
 is going indeed through a phase of intense activity, boosted primarily by large scale private initiatives such as the Breakthrough Listen program \citep{Enriquez2017,Isaacson2017,Price2019} and benefiting of significant advances in detector technologies.

The searches for bio- and technosignatures are two complementary strategies in the general quest of finding life elsewhere 
as they probe different remotely detectable byproducts of life.
Future searches for spectroscopic biosignatures will however target exoplanets up to only a few tens of light years from Earth \citep{Seager2018},
while current telescopes are potentially capable of detecting, for example, radio emissions from artificial sources of technological
level comparable to our own located well beyond 100 ly from the Earth \citep{Gray2020}. Moreover, 
searches for remotely detectable technosignatures probe a parameter search space so large that the absence of detection to date
does not warrant any firm conclusion about the existence of potentially detectable exocivilizations \citep{Wolfe1981,Tarter2001}.
This point has been recently emphasized by \citet{Wright2018} who compare the tiny fraction of the parameter space explored so far to the ratio of the volume of a small swimming pool to that of the Earth's oceans.

On the theoretical side, the prospects of technological life existing elsewhere and the probability of its detection are undetermined
as well. In this context, the famous Drake equation has traditionally inspired the search for electromagnetic (EM) technosignatures \citep{Drake1961,Drake1965}. 
In its compact form, the Drake equation equates the mean number of active emitters, $N_D$, with the product between the average rate of emergence of 
communicating civilizations, $\Gamma$, and the average longevity, $\bar{L}$, of the emission processes:
\begin{equation}
\label{Drake}
N_D=\Gamma\bar{L}.
\end{equation} 
In the original formulation of Eq.~\eqref{Drake}, $\Gamma$ was expressed as a product of probability factors, later grouped in astrophysical and 
biological/evolutionary probability events \citep{Prantzos2013}, though to be conducive to the emergence of technological life capable of releasing EM 
emissions. A vast literature has been devoted to the analysis of \eqref{Drake} \citep{Cirkovic2004,Maccone2010,Frank2016,Glade2016,Balbi2018} and
to the discussion of each entry of the original Drake equation thought, nowadays, 
only the astrophysical contributions to $\Gamma$ are known with some confidence \citep{Anchordoqui2018,Prantzos2019}.
Not surprisingly, lacking any empirical knowledge about the rate of abio- and technogenesis beyond Earth or the size of $\bar{L}$, 
estimates of $N_D$ span several orders of magnitude, even for the Milky Way galaxy alone \citep{Forgan2009}.

Although Eq. \eqref{Drake} is the most celebrated equation in the field of technosignatures, it is not the only vehicle 
to study their statistical properties. For example, of particular importance for assessing the probability of detection are the emission 
processes that cross our planet, as only those can be potentially detected \citep{Grimaldi2017}. Their average number, denoted $\bar{k}$, is thus 
a quantity that can be empirically estimated from observations, at least in principle.
In practice, however, the limited sensitivity of current telescopes and the aforementioned vastness of the parameter search space
permits only a probabilistic inference of the range of possible values of $\bar{k}$ compatible with observations \citep{Grimaldi2018,Flodin2019}.

Even if $\bar{k}$ is sometimes confused with the Drake number, it actually coincides with $N_D$ only by assuming
a constant birthrate of emissions that are either entirely isotropic, that is, radiating in all directions, or otherwise all directed 
towards our planet. In more general scenarios that contemplate anisotropic EM emissions, such as randomly directed beam-like 
signals or beacons sweeping across space, only the fraction of emissions directed towards the Earth can be potentially
detected, implying $N_D\geq \bar{k}$.

Another quantity of interest discussed here is $N_G$, the average total number of emission
processes present in the Galaxy. According to this definition, $N_G$ contains both the processes generated
by emitters that are still transmitting, whose average number is $N_D$, and those that come from emitters that are no longer active, 
but whose emissions still occupy the Galaxy (regardless of whether or not they intersect the Earth). 
In full generality, $N_G\geq N_D$, where the equality sign holds true either if there are no galactic emissions ($N_G=0$ and $N_D=0$) 
or if the only emissions present in the Galaxy come from emitters currently radiating.

The three quantities $N_G$, $N_D$, and $\bar{k}$ are the main statistical parameters characterizing the demography
of technosignatures in the Milky Way, from which other quantities and properties of interest can be derived. For example $\bar{k}/N_G$ gives the
fraction of galactic emissions intersecting Earth, while it can be demonstrated that $1-\exp(-\bar{k})$ yields the fraction of the galactic volume 
occupied by the emissions \citep{Grimaldi2017}. Furthermore, we see from the discussion above that the sequence of nested inequalities,
\begin{equation}
\label{inequality}
N_G\geq N_D\geq \bar{k},
\end{equation}
holds true for all types of EM emissions (isotropic, anisotropic) and for any combination of them, implying that
$\bar{k}$, the quantity that could be possibly estimated from observations, sets a lower limit to the population ($N_G$) of 
emissions filling the Galaxy. 

Here, we present a detailed study of $N_G$, $N_D$, and $\bar{k}$ to ascertain their dependence on the birthrate, the longevity, and 
the geometry of the emission processes.  We base our analysis on the presumption that the rate of technogenesis in the Milky Way has 
not changed significantly during the recent history of the Galaxy (a few million years) and that the population of artificial EM
sources can be described by a collection of statistically independent emitters. 

\begin{table*}
	\centering
	\caption{Legend of symbols used in the text and their meaning. The subscript $i$ refers to the different types of emissions considered here:
	 isotropic signals ($i=$ iso), random beams ($i=$ rb), and  rotating lighhouses ($i=$ lh).}
	\label{table1}
	\begin{tabular}{ll} 
		\hline
		symbol & meaning \\
		\hline
		$N_D^i$ & Drake's number of active emitters of type $i$\\
		$\bar{k}_i$ & average number of $i$-emissions crossing Earth at the present time\\
		$N_G^i$ & average number of $i$-emission intersecting the galactic plane\\
		$\Gamma_i$ & birthrate of emissions of type $i$\\
		$\bar{L}_i$ & mean longevity of emissions of type $i$\\
		$R_G$ & radius of the galactic plane ($\sim 60$ kly)\\
		$t_G=2R_G/c$ & travel time of a photon between two opposite edges of the galaxy ($\sim 10^5$ yr)\\
		$\alpha_0$ & angular aperture of a conical beam signal\\ 
		\hline
	\end{tabular}
\end{table*}

\section{Emission processes}
\label{processes}

We start by defining our model of emission processes generated
by artificial emitters in the Milky Way galaxy. We focus on the thin disk component of the Galaxy 
containing roughly $10^{10}$ potentially habitable planets within a radius $R_G\simeq 60$ kly from the galactic centre,
located at the origin of a cartesian reference frame with axes $x$, $y$, and $z$.
We approximate the thin disk by an effectively two-dimensional disk of radius $R_G$ on the $x$-$y$ plane.
We make the assumption that the artificial emitters are located at random sites $\mathbf{r}=(x,y)$ relative to the
galactic centre and that they are statistically independent of each other, meaning that their
birthrates and longevities  are random variables uncorrelated with $\mathbf{r}$.   

In the following, we shall employ the generic term ``emission process" to indicate an artificial EM radiation of any
wavelength and power spectrum that is emitted either continuously or not during a time duration
$L$. We shall however distinguish the emission processes according to their isotropy/anisotropy and to the geometry 
of the volume occupied by their radiations by assigning to them a distinct type or class labelled by the index $i$.
In particular, here we shall focus on three prototypical types of emissions: isotropic radiations ($i=$ iso), randomly 
directed narrow beams (denoted as ``random beams", $i=$ rb),
and narrow beams emitted by rotating lighthouses (denoted simply as ``lighthouses", $i=$ lh).
Furthermore, we shall assume that the radiations propagate unperturbed throughout the Galaxy at the speed of light $c$.
In so doing, we are neglecting scattering and absorption processes by the interstellar medium, which is a highly idealized 
setup meant to illustrate more clearly the effects of the emission geometries.

We allow the possibility that the different types of emission processes can have correspondingly different birthrates and longevities. 
For example, a continuous isotropic emission in the 
infrared could be resulting from the waste heat produced by a civilization exploiting the energy of its sun \citep{Dyson1960}, as for type II civilizations of the 
Kardashev scale \citep{Kardashev1964}. The corresponding emission birthrate would be
presumably lower, and the emission longevity longer, than that of a less technologically developed (or less energy harvesting) civilization targeting 
other planets with radio signals to just advertise its existence. We therefore 
introduce the rate of appearance per unit area for emitters of type $i$,
$\gamma_i(\mathbf{r},t)$, defined so that $\gamma_i(\mathbf{r},t)d\mathbf{r}dt$ gives the
expected number of $i$-emitters within an area element $d\mathbf{r}$ about $\mathbf{r}$ that started emitting within a time interval $dt$ centered at a
time $t$ before present. Likewise, we associate to all processes of type $i$ a common probability 
distribution function (PDF) of the longevity, denoted $\rho_i(L)$, such that $\int_0^\infty\!dL\,\rho_i(L)=1$ for each $i$.

The emitter rate of emergence vanishes for distances on the galactic disk larger than $R_\textrm{G}$, so that integrating $\gamma_i(\mathbf{r},t)$ 
over $\mathbf{r}$ gives the birthrate of the emission processes of type $i$ in the entire Galaxy:
\begin{equation}
\label{rate1}
\Gamma_i(t)=\int\!d\mathbf{r}\,\gamma_i(\mathbf{r},t).
\end{equation}
Finally, owing to the statistical independence of the emission processes, the sum of the birthrates of each type of emission gives the total
rate of appearance of all emission processes:
\begin{equation}
\label{rat2e}
\Gamma(t)=\sum_i \Gamma_i(t).
\end{equation}

\subsection{Average number of active emitters (Drake's $N_D$)}
\label{ND}

The Drake equation can be directly derived from considering the number of emitters that are currently transmitting \citep{Grimaldi2018a}.
To see this, we note that for any galactic emitter that started an emission process at a time $t$ in the past, the necessary condition that at present time the
emitter is still active is that the time elapsed since its birth is shorter than the emission longevity, that is, $t\leq L$.
The time integral of $\Gamma_i(t)$ from $t=0$ to $t=L$ gives therefore the expected number of emitters of type $i$ and
longevity $L$  that are still emitting. The average number of active $i$-emitters is obtained by marginalizing $L$ with respect 
to the PDF associated to the processes of type $i$:
\begin{equation}
\label{ND1}
N_D^i=\int_0^\infty\!dL\, \rho_i(L)\int_0^L\! dt\,\Gamma_i(t).
\end{equation}
We take the time-scale over which the rate $\Gamma_i(t)$ is expected to show appreciable variations to be much larger than $L$, even for longevity values 
distributed over several million years. In so doing, we are assuming that the emitter birthrates did not change significantly during the recent history of the Galaxy 
and can be taken constant in Eq.~\eqref{ND1}, leading to:
\begin{equation}
\label{ND2}
N_D^i=\Gamma_i\bar{L}_i,
\end{equation}
where $\bar{L}_i=\int_0^\infty\!dL\,\rho_i(L)L$ is the mean longevity of the emission processes of type $i$ (see Table \ref{table1} for a list of symbols used in 
this paper and their meaning). Equation \eqref{ND2} is the Drake
equation relative to signals of type $i$ under the steady-state hypothesis. By setting $x_i=\Gamma_i/\Gamma$ with $\sum_i x_i=1$, and defining 
\begin{equation}
\label{Lbar}
\bar{L}=\sum_i x_i\bar{L}_i
\end{equation}
as the longevity averaged over all types of emission processes, the sum of $N_D^i$ over all $i$'s leads to the usual Drake equation
in the compact form:
\begin{equation}
\label{ND3}
N_D=\Gamma \bar{L}.
\end{equation}
It is worth stressing that the steady-state hypothesis upon which the derivation of  Eq.~\eqref{ND3} rests would be less justifiable
if we were considering active emitters from a region extending over Giga light-years, as in this case the temporal 
dependence of the emission birthrates should be taken into account.

\subsection{Average number of emission processes at Earth ($\bar{k}$)}
\label{kbar}

While in deriving $N_D$ we only needed to count the number of active emitters without specifying the
characteristics of their emission processes, to calculate the mean number of emission processes intersecting Earth,
$\bar{k}$, we have to specify the conditions under which such intersections occur. First, we note
that for an emission process that started at a time $t$ in the past, the emitted EM radiations, traveling through
space at the speed of light $c$, fill at the present time a more or less extended region of space that depends on the
longevity and directionality of the emission process.
This region can be continuous, as in the case of an emission process (either isotropic or anisotropic) lasting a time $L$ without interruptions, 
or discontinuous as for an emitter having sent during $L$ a sequence of intermittent signals. 
In the latter case, if we assume that the train of signals is crossing Earth, there is a finite probability of the Earth not being illuminated
at a given instant of time \citep{Gray2020}, which may lead us to overlook this process in the calculation of $\bar{k}$.

Similar considerations apply also to intrinsically continuous emission processes that appear discontinuous or intermittent from the
Earth's viewpoint, as it is the case of a rotating beacon whose beamed signal crosses Earth periodically. Also, the signal intermittency 
may be due to variations in the emitted power with minimum flux at the receiver below the detection threshold \citep{Gray2020} or to scintillation
effects due to the interstellar medium \citep{Cordes1997}.

To avoid ambiguities in determining $\bar{k}$, we shall treat any intermittent (as seen from the Earth) emission of total longevity $L$ as an effectively 
continuous process lasting the same amount of duration time. Operatively, we could think of a periodic signal of period $T$ and duty cycle $w$ 
impinging upon the Earth during an observational time interval $\Delta t$. The condition $\Delta t/T\geq 1-w$ ensures that the ``on" phase of the emission
crosses Earth at least once during $\Delta t$, so that the process is ``detectable" with probability one and can be added to the list of processes
crossing Earth. We further note that among the requisites an intermittent signal should have to be recognized as a \textit{bona fide} 
technosignature, the recurrence of detection is one of the most important \citep{Forgan2019}.
  
\subsubsection{Isotropic emissions}
\label{iso}

Let us consider an emitter located at $\mathbf{r}$ that started emitting an isotropic process at a time $t$ in the past and for a duration $L$.
If the process is intrinsically continuous, at the present time the region of space filled by the EM waves
is a spherical concentric shell of outer radius $ct$ and thickness $cL$, centered on the emitter position $\mathbf{r}$. 
In the case of a intermittent isotropic process of period $T$ and duty cycle $w$, this region encompasses a sequence of nested concentric 
spherical shells, each of thickness $cwT$, of consecutive outer radii differing by $cT$. 
Conforming to the above prescription for
intermittent emissions, we ignore the internal structure of the encompassing shell by treating it as an effectively continuous spherical shell of width $cL$.

The condition that the shell intersects the Earth is fulfilled by the requirement \citep{Grimaldi2017,Balbi2018}
\begin{equation}
\label{indi1}
ct-cL\leq\vert\mathbf{r}-\mathbf{r}_\textrm{E}\vert\leq ct,
\end{equation}
where $\mathbf{r}_\textrm{E}$ is the vector position of the Earth. 
If $\gamma_\textrm{iso}(\mathbf{r},t)$ (where the subscript ``iso" stands for isotropic) is the process
birthrate per unit volume and $\rho_\textrm{iso}(L)$ is the PDF of the longevity, the average number of
spherical shells at Earth is obtained by marginalizing the condition \eqref{indi1} over $\mathbf{r}$, $L$, and $t$:
\begin{equation}
\label{k1}
\bar{k}_\textrm{iso}
=\int\! dL\,\rho_\textrm{iso}(L)\int\! d\mathbf{r}\int_{\vert\mathbf{r}-\mathbf{r}_\textrm{E}\vert/c}^{\vert\mathbf{r}-\mathbf{r}_\textrm{E}\vert/c+L}\! \!\! dt\,\gamma_\textrm{iso}(\mathbf{r},t).
\end{equation}
As done in Sec.~\ref{ND}, we neglect the temporal dependence of the birthrate, $\gamma_\textrm{iso}(\mathbf{r},t)\simeq \gamma_\textrm{iso}(\mathbf{r})$, so that Eq.~\eqref{k1} reduces to: 
\begin{equation}
\label{k2}
\bar{k}_\textrm{iso}=\int\! dL\,\rho_\textrm{iso}(L) L\int\! d\mathbf{r}\,\gamma_\textrm{iso}(\mathbf{r})=\Gamma_\textrm{iso}\bar{L}_\textrm{iso},
\end{equation}
which, as anticipated in Sec.~\ref{ND}, coincides with the Drake number $N_D^\textrm{iso}$ relative to isotropic emission processes.
Note that since we have taken a time independent birthrate, $\mathbf{r}_E$ has dropped off Eq.~\eqref{k2}, meaning that $\bar{k}_\textrm{iso}$ actually 
gives the mean number of emissions crossing any given point in the Galaxy. This holds true also for other types of emission processes as long as the
corresponding birthrates do not depend on $t$. 

\subsubsection{Anisotropic emissions: random beams and lighthouses}
\label{anisok}
In the case of anisotropic signals, the region of space filled by the EM radiation does not cover all directions and, therefore, only the fraction
of signals that are directed towards the Earth can contribute to $\bar{k}$. For example, a prototypical anisotropic signal often discussed in the 
literature is that of a conical beam of opening angle $\alpha_0$ pointing to a given direction over the full lifetime of the emission process.
As shown in Appendix \ref{kbeams}, if such beamed signals are generated with a constant birthrate and their orientation is distributed uniformly over the 
unit sphere (3D case), the average number of random beams (rb) intersecting Earth will be proportional to the solid angle subtended by the beams, that is: 
\begin{equation}
\label{rb01}
\begin{array}{lll}
\bar{k}_\textrm{rb}=\dfrac{\langle\alpha_0^2\rangle}{16}\Gamma_\textrm{rb}\bar{L}_\textrm{rb}, & \textrm{(3D random beams)}
\end{array}
\end{equation}
where $\langle\cdots\rangle$ denotes an average over the beam apertures (assumed to be narrow), $\bar{L}_\textrm{rb}$ is the average
longevity of the beams, and $\Gamma_\textrm{rb}$ is their birthrate. We see therefore that contrary to the case of isotropic emission processes, the mean 
number of beams crossing Earth can be many orders of magnitude smaller than the corresponding Drake's number 
$N_D^\textrm{rb}=\Gamma_\textrm{rb}\bar{L}_\textrm{rb}$. 
Taking for example beam apertures comparable to that of the Arecibo radar ($\sim 2\arcmin\simeq 6\times 10^{-4}$ rad)
Eq.~\ref{rb01} yields $\bar{k}_\textrm{rb}/N_D^\textrm{rb}\sim 2\times 10^{-8}$, and 
even smaller values of $\bar{k}_\textrm{rb}/N_D^\textrm{rb}$ are obtained by assuming optical or infrared laser emissions of apertures under 
an arcsecond \citep{Howard2004,Tellis2017}. 

Instead of pointing towards random directions in space, another hypothetical scenario is that in which the emitters generate beams directed preferably 
along the galactic plane in order to enhance the probability of being detected by other civilizations. In this two dimensional (2D) case, the mean number of beams illuminating the 
Earth becomes (see \ref{kbeams}):
\begin{equation}
\label{rb02}
\begin{array}{ll}
\bar{k}_\textrm{rb}=\dfrac{\langle\alpha_0\rangle}{2\pi}\Gamma_\textrm{rb}\bar{L}_\textrm{rb}, & \textrm{(2D random beams)}
\end{array}
\end{equation}
so that $\bar{k}_\textrm{rb}/N_D^\textrm{rb}\sim 10^{-4}$ for $\langle\alpha_0\rangle$ comparable to that of the Arecibo radar. 

Another type of anisotropic signal is that of a narrow beam emitted by a rotating source, like a lighthouse rotating with constant angular velocity. 
This kind of process generates a continuous, radiation-filled spiralling beam 
revolving around the emitter and expanding at the speed of light. The spiral cross section 
increases with the beam aperture $\alpha_0$ and the distance from the source. Even if this kind of process is intrinsically continuous, 
an expanding spiral impinging upon Earth will be perceived as a periodic sequence of signals.
For example, if the spin axis is perpendicular to the galactic plane, the spiral generated by
a rotating conical beam of angle aperture $\alpha_0$ will periodically cross Earth's line-of-sight with duty 
cycle $\alpha_0/2\pi$.

\begin{figure}
	\begin{center}
			\includegraphics[width=\columnwidth,clip=true]{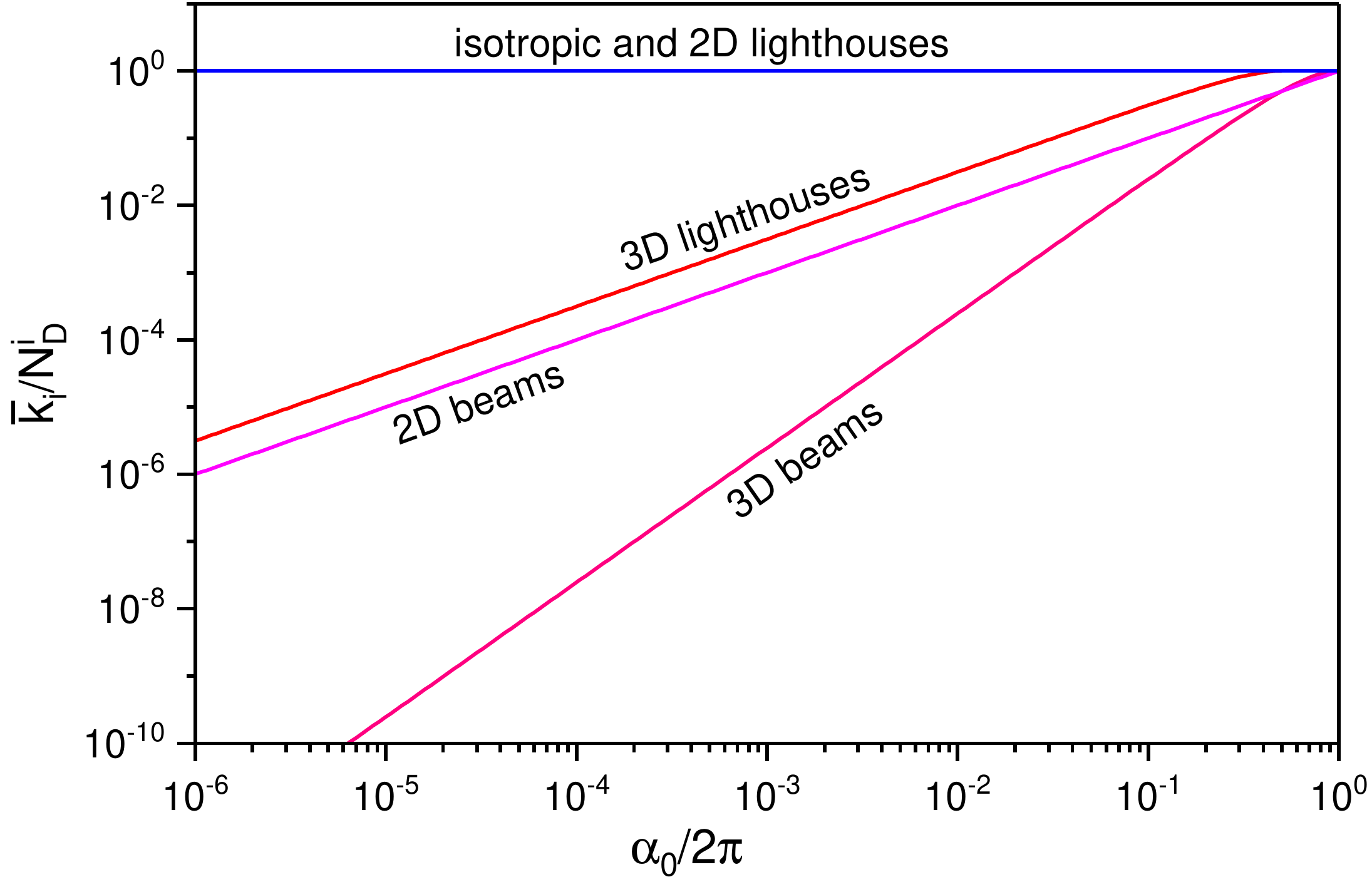}
		\caption{Average number of emissions of type $i$ crossing Earth, $\bar{k}_i$, over the corresponding Drake number $N_D^i$ plotted as a function 
		of the beam aperture $\alpha_0$ for the different types $i$ of processes considered in this work.}\label{fig1}
	\end{center}
\end{figure}

In a manner similar to what we have done for the case of intrinsically discontinuous signals, we discard the signal intermittency perceived at Earth 
by introducing an effective volume encompassing the spiralling beam, whose construction is detailed in the Appendix \ref{klighthouses}. As long as the rotation axes
are oriented uniformly over the unit sphere (3D case), we find that the mean number of lighthouse (lh) signals crossing Earth reduces to:
\begin{equation}
	\label{lh01}
	\begin{array}{cc}
	\bar{k}_\textrm{lh}=\dfrac{\displaystyle\langle\alpha_0\rangle}{\displaystyle 2}\Gamma_\textrm{lh}\bar{L}_\textrm{lh}, &\textrm{(3D lighthouses)}
	\end{array}
\end{equation}
while when the spin axes are perpendicular to the galactic plane (2D case), $\bar{k}_\textrm{lh}$ becomes:
\begin{equation}
	\label{lh02}
	\begin{array}{lr}
	\bar{k}_\textrm{lh}=\Gamma_\textrm{lh}\bar{L}_\textrm{lh}, & \textrm{(2D lighthouses)}.
	\end{array}
\end{equation}
The signals generated by lighthouses have therefore much larger values of $\bar{k}$ than those of random beams with comparable birthrates and 
longevities, as shown in Fig.~\ref{fig1}. In particular, from Eqs.~\eqref{rb01}-\eqref{lh02} we see that $\bar{k}_\textrm{rb}/\bar{k}_\textrm{lh}$ scales as 
$\sim\langle\alpha_0\rangle N_D^\textrm{rb}/N_D^\textrm{lh}$ for 2D or 3D anisotropic signals of similar apertures, meaning for example that
over $\sim 10^3$ active Arecibo-like beams must thus be added to each active lighthouse of comparable 
$\langle\alpha_0\rangle$ to have analogous values of $\bar{k}$. Furthermore, $\bar{k}_\textrm{lh}$ for the 2D case turns out to be independent 
of $\alpha_0$, as for isotropic processes. 
This is an interesting result, implying that an observer on Earth has the same chances of being illuminated by a 2D lighthouse as 
by a (pulsed) isotropic signal if the two have equal Drake's numbers, Fig.~\ref{fig1}.

\subsection{Average number of emission processes intersecting the Galaxy ($N_G$)}
\label{NG}

So far, the only temporal variable required to calculating $N_D^i$ and $\bar{k}_i$ has been the signal longevity $L$. 
In deriving $N_G$, we shall introduces an additional time-scale,
 \begin{equation}
 \label{tG}
t_G=\frac{2R_G}{c},
 \end{equation} 
defined as the time required by a photon to travel, unperturbed, across two opposite edges of the Milky Way ($\approx 10^5$ yr).  
Contrary to $L$ and $\Gamma_i$, $t_G$ is an astrophysical quantity, specific to our Galaxy, that is independent of any assumption about the 
existence and/or the properties of the artificial emissions.

\begin{figure}
	\begin{center}
		\includegraphics[scale=0.7,clip=true]{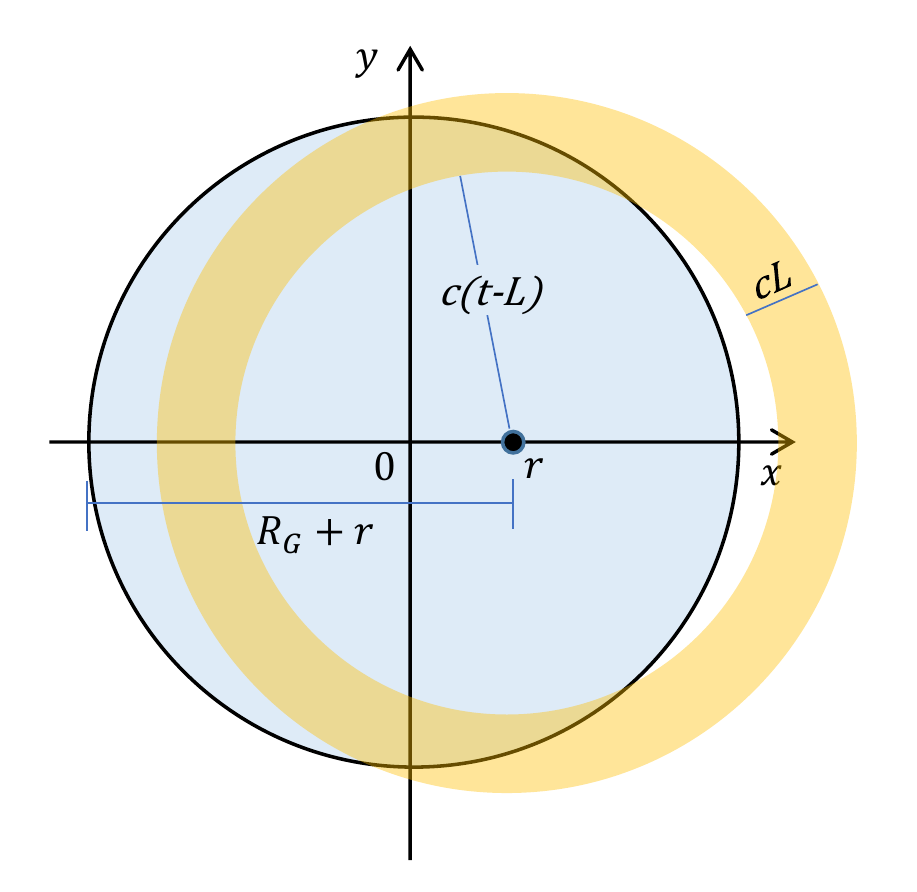}
		\caption{Top view of the galactic disk of radius $R_G$ with superimposed areas covered by emissions of different geometries originated by a source
			located at $(r,0)$. (a): the intersection with the plane $x$-$y$ of the spherical shell generated by an isotropic emission forms an annulus of inner radius $c(t-L)$,
			which intersects the galactic disk if $c(t-L)\leq R_G+r$.}\label{fig2}
	\end{center}
\end{figure}

\subsubsection{Isotropic emissions}

The necessary condition for an isotropic signal intersecting the Galaxy is that there is a non-null intersection between the spherical shell
and the galactic disk. As shown in Fig.~\ref{fig2}, this is fulfilled by requiring the inner radius of the shell to be smaller than the maximum distance of the emitter from the
edge of the galactic disk, that is, $c(t-L)\leq r+R_G$. The number of isotropic emissions intersecting the Galaxy is therefore:
\begin{align}
\label{NGiso1}
N_G^\textrm{iso}&=\int\! dL\rho_\textrm{iso}(L)\int\! d\mathbf{r}\!\int_0^{(r+R_G)/c+L}\! dt\,\gamma_\textrm{iso}(\mathbf{r})\nonumber \\
&=\Gamma_\textrm{iso}\bar{L}_\textrm{iso}+\frac{1}{c}\int\! d\mathbf{r}\gamma_\textrm{iso}(\mathbf{r})(r+R_G).
\end{align}
Using a birthrate that is uniform over the entire galactic disk, $\gamma_\textrm{iso}(\mathbf{r})=\theta(R_G-r)\Gamma_\textrm{iso}/\pi R_G^2$, the above 
expression reduces to:
\begin{equation}
\label{NGiso2}
N_G^\textrm{iso}=\Gamma_\textrm{iso}\left(\bar{L}_\textrm{iso}+\frac{5}{6}t_G\right),
\end{equation}
where $t_G$ is the time-scale given in Eq.~\eqref{tG}. Other functional forms of $\gamma_\textrm{iso}(\mathbf{r})$ affects only the prefactor of $t_G$.
For example, taking $\gamma_\textrm{iso}(\mathbf{r})\propto \theta(R_G-r)\exp(-r/r_s)$ with $r_s=8.15$ kly \citep{Grimaldi2018}, the numerical 
factor $5/6$ ($\simeq 0.833$) in \eqref{NGiso2} becomes $\simeq 0.646$.

An interesting feature of Eq.~\eqref{NGiso2} is that using \eqref{k2} we can replace $\Gamma_\textrm{iso}$ by $\bar{k}_\textrm{iso}/\bar{L}_\textrm{iso}$, yielding:
\begin{equation}
\label{NGiso3}
N_G^\textrm{iso}=\bar{k}_\textrm{iso}\left(1+\frac{5}{6}\frac{t_G}{\bar{L}_\textrm{iso}}\right),
\end{equation}
so that for $\bar{L}_\textrm{iso}\ll t_G\approx 10^5$ yr the expected number of the emissions intersecting the Galaxy can be much larger than that 
of the emissions crossing Earth. For example, even if $\bar{k}_\textrm{iso}$ is only $\approx  0.1$ and $\bar{L}_\textrm{iso}\approx 10^2$ yr,
$N_G^\textrm{iso}$ is nevertheless of the order $10^2$. As we shall see below, the directionality of the signals can amplify even more the 
difference between $\bar{k}_i$ and $N_G^i$. 

\subsubsection{Anisotropic emissions: random beams and lighthouses}
\label{anisoNG}
The calculations of the average number of random beams present in the Galaxy, $N_G^\textrm{rb}$, and the one relative to the lighthouse signals, $N_G^\textrm{lh}$, 
are detailed in the Appendixes \ref{NGbeams} and \ref{NGlighthouses}, respectively. Here we report only the final expressions obtained under the assumption 
of small opening angles and a spatially uniform birthrate of the emitters:
\begin{equation}
	\label{NGrb}
	N_G^\textrm{rb}=\left\{
	\begin{array}{lr}
		\Gamma_\textrm{rb}\left(\bar{L}_\textrm{rb}+\dfrac{\langle\alpha_0\rangle}{2}\dfrac{4}{3\pi}t_G\right), & \textrm{3D random  beams} \\[8pt]
		\Gamma_\textrm{rb}\left(\bar{L}_\textrm{rb}+\dfrac{4}{3\pi}t_G\right), & \textrm{2D random beams},
	\end{array}\right.
\end{equation}
\begin{equation}
	\label{NGsp}
	N_G^\textrm{lh}=\left\{
	\begin{array}{lr}
		\Gamma_\textrm{lh}\left(\bar{L}_\textrm{lh}+\dfrac{2}{\pi}t_G\right), & \textrm{3D lighthouses} \\[8pt]
		\Gamma_\textrm{lh}\left(\bar{L}_\textrm{lh}+\dfrac{5}{6}t_G\right), & \textrm{2D lighthouses}.
	\end{array}\right.
\end{equation}
The relevant result of these calculations is that for all but one case (that is, the 3D random beams) the mean number of anisotropic emissions
intersecting the galactic plane is independent of the angular aperture, and is therefore comparable to that obtained for the case of isotropic 
emissions with similar birthrates and longevities.

\section{Discussion}
\label{discu}

\begin{table}
	\centering
	\caption{Analytic expressions of the average number of emissions intersecting the Galaxy, $N_G$, and the expected
	number of processes at Earth, $\bar{k}$, for the different types of emissions considered in this article. $\Gamma_i$ is the birthrate of emissions 
	of type $i$ (with $i=$ iso for isotropic signals, $i=$ rb for random beams, and $i=$ lh for rotating lighthouses) 
	and $\bar{L}_i$ is the corresponding mean longevity. $\alpha_0$ denotes the beam aperture. For all cases, the corresponding 
	Drake's number is $N_D^i=\Gamma_i\bar{L}_i$. }
	\label{table2}
	\begin{tabular}{lll} 
		\hline
		type & $N_G$ & $\bar{k}$\\
		\hline
		isotropic & $\Gamma_\textrm{iso}\!\left(\bar{L}_\textrm{iso}+\frac{5}{6}t_G\right)$  & $\Gamma_\textrm{iso}\bar{L}_\textrm{iso}$\\[4pt] 
		3D random beams & $\Gamma_\textrm{rb}\!\left(\bar{L}_\textrm{rb}+\frac{\langle\alpha_0\rangle}{2}\frac{4}{3\pi}t_G\right)$ & 
		 $\frac{\langle\alpha_0^2\rangle}{16}\Gamma_\textrm{rb}\bar{L}_\textrm{rb}$ \\[4pt]
		2D random beams & $\Gamma_\textrm{rb}\!\left(\bar{L}_\textrm{rb}+\frac{4}{3\pi}t_G\right)$ &
		$\frac{\langle\alpha_0\rangle}{ 2\pi}\Gamma_\textrm{rb}\bar{L}_\textrm{rb}$ \\[4pt]
		3D lighthouses & $\Gamma_\textrm{lh}\!\left(\bar{L}_\textrm{lh}+\frac{2}{\pi}t_G\right)$  
		& $\frac{\langle\alpha_0\rangle}{2}\Gamma_\textrm{lh}\bar{L}_\textrm{lh}$ \\[4pt]
		2D lighthouses & $\Gamma_\textrm{lh}\!\left(\bar{L}_\textrm{lh}+\frac{5}{6}t_G\right)$  
		& $\Gamma_\textrm{lh}\bar{L}_\textrm{lh}$ \\
		\hline
	\end{tabular}
\end{table}

Table \ref{table2} summarizes the analytic expressions of $N_G^i$ and $\bar{k}_i$ 
derived in the previous section. For each type of emission process, the Drake number $N_D^i=\Gamma_i\bar{L}_i$ is the only quantity 
that does not depend on the geometry of the emission process and we shall therefore focus our discussion primarily on $\bar{k}_i$ and $N_G^i$.

\begin{figure}
		\includegraphics[scale=0.4,clip=true]{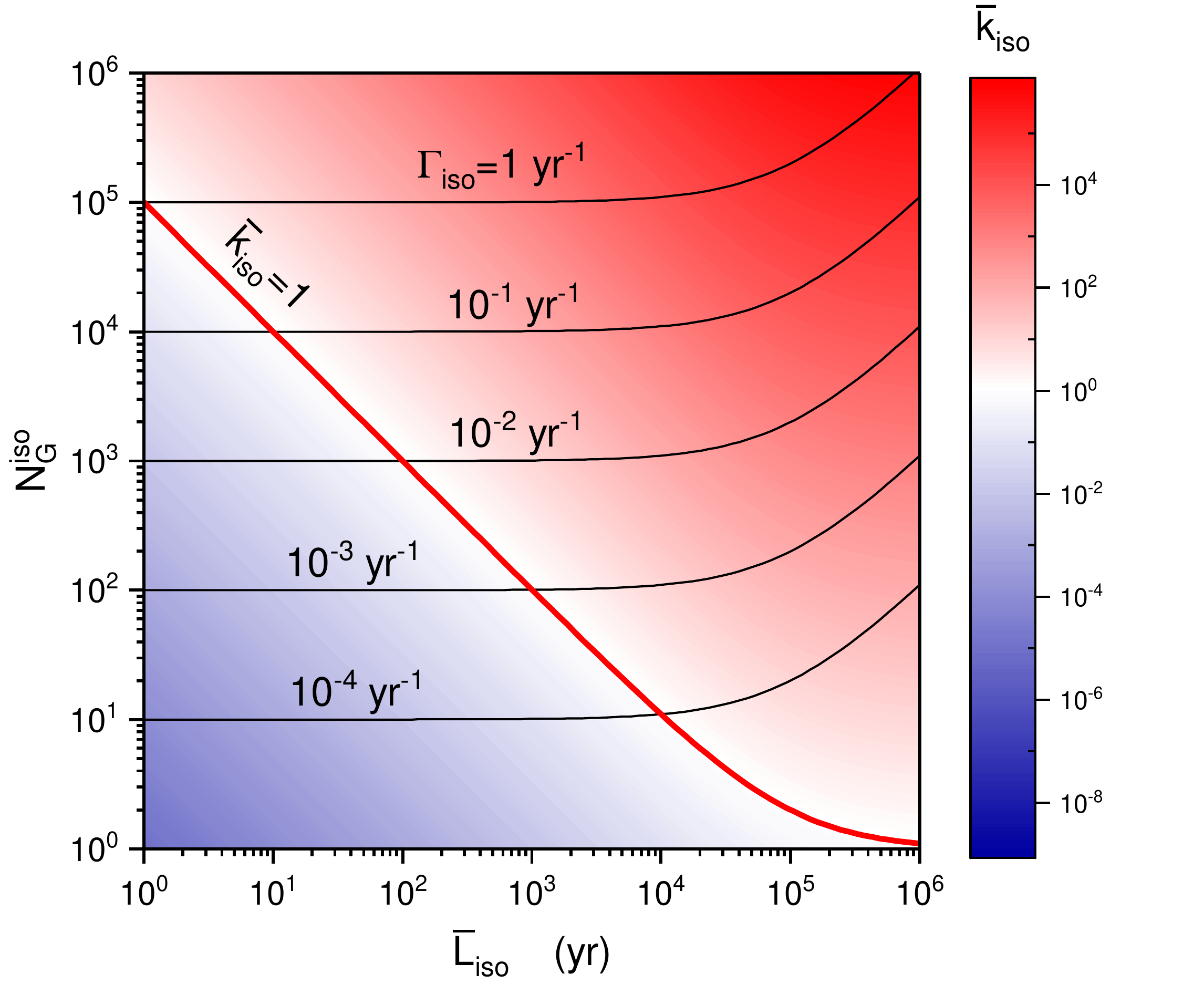}
		\caption{Expected number of isotropic emission processes at Earth, $\bar{k}_\textrm{iso}$, as a function of the mean signal longevity
		$\bar{L}_\textrm{iso}$ and the average number $N_G^\textrm{iso}$ of isotropic emissions intersecting the Galaxy. The red solid line 
		indicates the special value $\bar{k}_\textrm{iso}=1$, while the black solid lines show $N_G^\textrm{iso}$ calculated for different values 
		of the emission birthrate $\Gamma_\textrm{iso}$.}\label{fig3}
\end{figure}

\begin{figure*}
	\begin{center}
			\includegraphics[scale=0.4,clip=true]{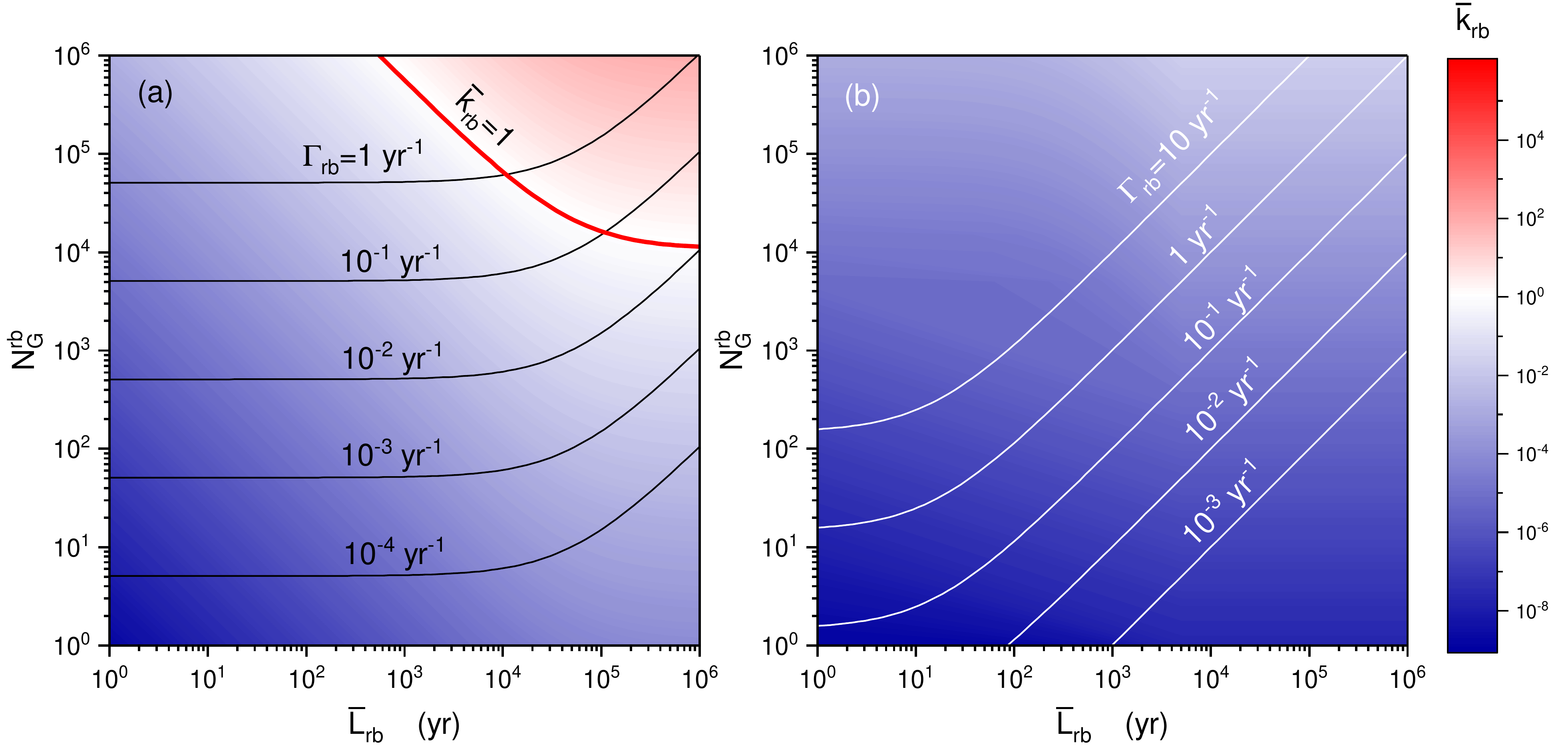}
		\caption{Expected number of random narrow beams crossing Earth, $\bar{k}_\textrm{rb}$, as a function of their mean longevity
		$\bar{L}_\textrm{rb}$ and the average number $N_G^\textrm{rb}$ of beams intersecting the Galaxy. In panel (a) the orientations of the beam 
		axes are distributed uniformly along the galactic disk (2D case), while in panel (b) they are distributed uniformly over the 3D space. The
		beam aperture is fixed at $2\arcmin$, corresponding to $\alpha_0=6\times 10^{-4}$ rad.
		The red solid lines indicate the special value $\bar{k}_\textrm{rb}=1$, while the black solid lines show $N_G^\textrm{rb}$ calculated for 
		different values of the emission birthrate $\Gamma_\textrm{rb}$.}\label{fig4}
	\end{center}
\end{figure*}

\begin{figure*}
	\begin{center}
			\includegraphics[scale=0.4,clip=true]{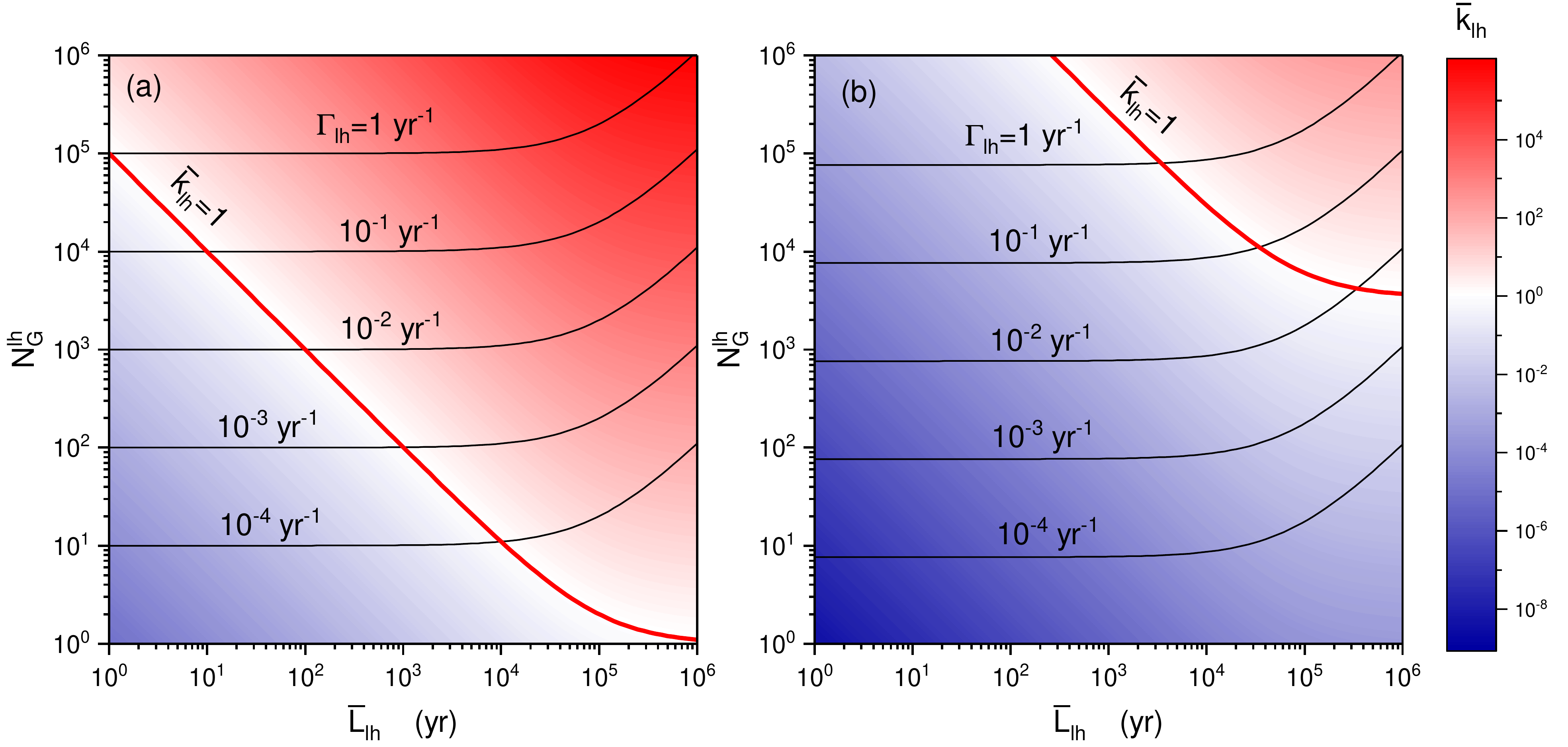}
		\caption{Expected number of emissions from rotating beacons crossing Earth, $\bar{k}_\textrm{lh}$, as a function of their mean longevity
		$\bar{L}_\textrm{lh}$ and the average galactic population $N_G^\textrm{lh}$. In panel (a) the orientations of the spin
		axes are kept perpendicular to the galactic plane (2D case), while in panel (b) they are distributed uniformly over the 3D space. 
		The red solid lines indicate the special value $\bar{k}_\textrm{lh}=1$, while the black solid lines show $N_G^\textrm{lh}$ calculated for 
		different values of the emission birthrate $\Gamma_\textrm{lh}$.}\label{fig5}
	\end{center}
\end{figure*}

Figures \ref{fig3}-\ref{fig5} show $\bar{k}_i$ as a function of the mean signal longevity 
$\bar{L}_i$ and the population of signals in the Galaxy $N_G^i$, with $i=\textrm{iso}$ (isotropic, Fig.\ref{fig3}), $i=\textrm{rb}$  (random beams,Fig.\ref{fig4}), and
$i=\textrm{lh}$ (lighthouses, Fig.\ref{fig5}). 
The results have been obtained by taking $R_G=60$ kly for the galactic radius, corresponding to $t_G=2R_G/c=1.2\times 10^{5}$ yr. 
The red solid lines demarcate the boundary between
$\bar{k}_i >1$ (red colour scale) and $\bar{k}_i <1$ (blue colour scale), while the black solid lines indicate 
$N_G^i$ calculated for constant values of the emission birthrate $\Gamma_i$. The results shown in Figs.~\ref{fig4} and \ref{fig5} 
have been obtained assuming $\langle\alpha_0^2\rangle\simeq \langle\alpha_0\rangle^2$ and an average beam aperture 
of $2\arcmin$, corresponding to $\langle\alpha_0\rangle\simeq 6\times 10^{-4}$ rad. Results for different beam apertures can be easily 
obtained using the expressions in Table \ref{table2}. 

A first interesting feature is the behaviour of the galactic population of technosignatures, $N_G^i$, as a function of $\bar{L}_i$ for 
fixed $\Gamma_i$ (black solid lines). While $N_G^i$ increases proportionally to the signal longevity for $\bar{L}_i\gtrsim  t_G$, reaching 
asymptotically the corresponding Drake's number $N_D^i=\bar{L}_i\Gamma_i$, for $\bar{L}_i\lesssim t_G\sim 10^5$ yr it reduces  to
\begin{equation}
\label{NGsmallL}
N_G^i\simeq\Gamma_i t_G\simeq \Gamma_i\times(10^5 \textrm{yr}),
\end{equation}
for all types of emission processes with the exception of random beams in 3D. In this case 
$N_G^\textrm{rb}$ scales as $\langle\alpha_0\rangle \Gamma_\textrm{rb} t_G$ for $\bar{L}_\textrm{rb}\lesssim \langle\alpha_0\rangle t_G$.


Equation \eqref{NGsmallL} is remarkable because it prescripts the galactic population of emission processes to be proportional to only the birthrate 
$\Gamma_i$, regardless of the signal longevity as long as it is assumed $\bar{L}_i$ to be less than $\sim 10^5$ yr. This is an advantage compared to the Drake's 
number $N_D^i$, where in addition to $\Gamma_i$ the longevity of the signals is a further object of speculations. For a wide range of $\bar{L}_i$ values,
we can thus conjecture about the size of $N_G^i$ by reasoning only in terms of the signal birthrate. To this end it is instructive to compare $\Gamma_i$
with the rate of formation of habitable planets in the Milky Way, $\Gamma_\textrm{P}$, whose estimates place it in the range $0.01$-$0.1$ planet 
per year \citep{Behroozi2015,Zackrisson2016,Gobat2016,Anchordoqui2018}. 

Let us first make the hypothesis that each habitable planet can be the potential source of no more than one artificial emission. This would correspond
to $\Gamma_\textrm{P}$ being a theoretical upper limit of $\Gamma_i$.  Under this assumption, the resulting galactic population of
both isotropic emissions (Fig.~\ref{fig3}) and rotating beacons (Fig.~\ref{fig5}) would be bounded from above by $\max (N_G^i)\sim 10^3$-$10^4$,
or somewhat less for 2D random beams of Fig.~\ref{fig4}(a),
which is essentially the number of habitable planets being formed during a timespan of order $t_G\sim 10^5$ yr.

Such an upper limit of $N_G$ entails a corresponding lower bound on the average distance $a_E$ between the emitters. Indeed, since within
our working assumption $N_G$ corresponds to the number of emitters releasing the emissions, their number density can be expressed as
$\rho_E=N_G/\pi R_G^2$. This allows us to find from $\pi a_E^2\rho_E\sim 1$ that $a_E\sim R_G/\sqrt{N_G}$, thereby implying 
 that the lower bound on $a_E$ is of the order $10^3$ ly for $\bar{L}\lesssim t_G$.
 Following the same reasoning, we see that the typical relative distance between emitters estimated by the Drake equation, 
 $a_D\sim R_G/\sqrt{N_D}$, scales for $\bar{L}\lesssim t_G$ as $a_D\sim a_E\sqrt{t_G/\bar{L}}$. The difference between $a_G$ and $a_D$ stems from the
fact that the Drake's number gives the average population of active emitters, which are only a fraction of all $N_G$ emitters whose signals are present 
in the Galaxy. For example, while  assuming $\Gamma_i\sim\Gamma_P$ and $\bar{L}\lesssim 10$ yr gives $N_G\sim 10^3$ and $a_E\sim 10^3$ ly,  
the Drake equation yields $N_D\lesssim 1$ and a value of $a_D$ comparable to or larger than the diameter of the Galaxy, meaning that in this case out of 
$\sim 10^3$ galactic emissions essentially none comes from currently active emitters.

As we have seen in the previous section, in addition to $N_D$ the longevity $\bar{L}$ constraints also the number of the emission
processes crossing our planet, which is further affected by the directionality of the signals (Fig.~\ref{fig2}).
Assuming therefore a large number of galactic emissions does not automatically imply 
similarly large values of $\bar{k}_i$. For example, even taking $\Gamma_i\sim 0.1$ yr$^{-1}$ (that is, $N_G^i\simeq 10^4$ for $\bar{L}_i<t_G$), the 
expected maximum value of $\bar{k}_i$ 
ranges from $\sim 10^3$ for $\bar{L}_i=10^4$ yr down to $\sim 0.1$ for $\bar{L}_i=1$ yr in the case of isotropic emissions (Fig.~\ref{fig3}) and 2D lighthouses
[Fig.~\ref{fig5}(a)]. Within the same range of signal longevities, the upper bound on $\bar{k}_i$ of beams in 2D and rotating beacons in 3D drops to only  
$\sim 10^{-5}$-$10^{-1}$, Figs.~\ref{fig4}(a) and \ref{fig5}(b).

A special situation is represented by a collection of beamed signals with axis orientations distributed in the 3D space [Fig.~\ref{fig4}(b)]. In this case, $N_G^\textrm{rb}$
becomes independent of the signal longevity only when $\bar{L}_\textrm{rb}\lesssim \langle\alpha_0\rangle t_G$, which for $\langle\alpha_0\rangle\sim 2\arcmin$
represents lifetimes smaller than $\sim 10$ yr. In this limit, $N_G^\textrm{rb}\lesssim 1$ for $\Gamma_\textrm{rb}\sim 0.1$ yr$^{-1}$ and the resulting
$\bar{k}_\textrm{rb}$ is upper bounded by a negligible $\sim 5\times 10^{-8}$. Values of $N_G$ of the order of $10^5$ can nevertheless be reached for 3D beams
lasting at least $1$ Myr, but even in this case $\bar{k}_\textrm{rb}$ is only $\sim 2\times 10^{-3}$. 

Let us pause one moment to consider the implications of assuming each habitable planet being the potential source of at most one
emission process. As shown above, this hypothesis entails an upper bound of $N_G^i$ of the order $10^3$-$10^4$, implying therefore the possibility of 
technogenesis arising on \textit{each} habitable planet during the last $\sim 10^5$ years. 
This exceeds by far the most optimistic stances, as such an assumption would imply not only a non-zero probability that abiogenesis is 
ubiquitous in the Milky Way, but also that intelligence and
technology are inevitable outcomes of the evolutionary path of life on each inhabited planet. As long as a one-to-one correspondence between
emission processes and planets is maintained, an upper bound on $\Gamma_i$ (and so on $N_G^i$) should be more reasonably placed to
much lower values than $\Gamma_P$, leading to $\max(N_G^i)\ll 10^3$-$10^4$ and to correspondingly small values of $\bar{k}_i$.

Our model, however, does not distinguish whether the emission events have occurred once or multiple times within $t_G$ on a given planet, nor does it rule out the
possibility of the emitters far outnumbering the planets in which technology arose, as for example self-replicating robotized lighthouses swarming 
in the free space. Within such scenarios, $\Gamma_i$ could thus be larger than 
the rate of emergence of technological civilizations capable of releasing technosignatures and perhaps even
comparable to, or in excess of, $\Gamma_P$. The plausibility of a galactic population of $\sim 10^4$ short lived 
(i.e., $\bar{L}_i\lesssim t_G$) emissions should however be weighed against the requirement that all these emissions must have been
released during the last $\sim 10^5$ years in order to fill the galaxy. 

Of course, it is still possible to have significantly large values of $N_G^i$ and $\bar{k}_i$ even for relatively low birthrates if the mean longevity
is so long to prevail over the small values of $\Gamma_i$. For example, signals emitted from isotropic sources or 2D lighthouses with a rate 
as small as $\sim 10^{-5}$ yr$^{-1}$ would bring values of $N_G^i$ and $\bar{k}$ larger $\sim 100$ if their longevities exceeded 
$\sim 10$ Myr. Similar values of $N_G^i$ are obtained for 2D beams and 3D lighthouses with $\Gamma_i\sim 10^{-5}$ yr$^{-1}$ and
$\bar{L}_i\sim 10$ Myr, but the reduced solid angle for $\langle\alpha_0\rangle\sim 2\arcmin$ makes $\bar{k}_i$ as small as $\sim 10^{-2}$
(which drops to $\sim 10^{-8}$ in the case of 3D random beams).

So far, we have discussed each type of emission processes separately to study the effect of $\Gamma_i$, $\bar{L}_i$ and of
the signal directionality on $N_G^i$ and $\bar{k}_i$. However, in the most general case, different types of processes
may be present simultaneously in the Galaxy and the contribution of each $i$-process to the total $N_G$ and $\bar{k}$
depends on the respective occurrence frequency.  
To see this, we note that the different expressions of 
$N_G^i$ and $\bar{k}_i$ given in Table \ref{table2} have the form $N_G^i=\Gamma_i(\bar{L}_i+u_i t_G)$ and $\bar{k}_i=v_i\Gamma_i\bar{L}_i$,
where $u_i$ and $v_i$ are the dimensionless factors taking account the geometry and the directionality of the signals. 
Owing to the assumed statistical independence of the emitters, the quantities $N_D$, $N_G$, and $\bar{k}$ are 
simple linear combinations of the different types of processes. We can thus write:
\begin{align}
\label{NGtot1}
N_G&=\sum_i \Gamma_i(\bar{L}_i+u_it_G)=\Gamma\bar{L}+\Gamma t_G\sum_i x_i u_i, \\
\label{ktot1}
\bar{k}&=\sum_i v_i\Gamma_i\bar{L}_i=\Gamma\sum_ix_iv_i\bar{L}_i,
\end{align}
where, as done in Sec.\ref{ND}, $x_i=\Gamma_i/\Gamma$ and $\bar{L}=\sum_i x_i\bar{L}_i$. 
Considerations similar to those discussed in the previous section apply therefore also to the more general case. In particular, as seen
from Eq.~\eqref{NGtot1}, as long as the total signal longevity $\bar{L}$ is smaller than $\sim t_G\sim 10^5$ yr, the total number of processes intersecting the 
Galaxy results to be proportional to $\Gamma t_G\sim (10^5\,\textrm{yr})\times\Gamma$, regardless of $\bar{L}$. Speculations about
the abundance of short-lived ($\bar{L}\lesssim 10^5$ yr) emissions in the Galaxy can thus be framed in terms of possible upper bounds on the 
total birthrate $\Gamma$.

From Eq.~\eqref{ktot1} we see that the contribution of each type of emission to the total number of processes crossing Earth
strongly depends on the relative abundance of signal types and their longevities. As shown in Figs.~\ref{fig3}-\ref{fig5}, the contribution 
to $\bar{k}$ of isotropic processes and lighthouses in 2D would likely dominate over other types of emissions of similar birthrates. 
For example, assuming that the fraction of rotating beacons sweeping the galactic plane is comparable to that of 3D beamed 
emissions, $x_\textrm{lh}\sim x_\textrm{rb}$, 
the two would contribute equally to $\bar{k}$ only if the mean longevity of the 3D beams is about $16/\langle\alpha_0^2\rangle$ times larger than 
that of the rotating beacons. For beam apertures of $2\arcmin$ this corresponds to a factor $\sim 10^7$, so for a given fraction of 
2D lighthouses lasting in average $10$ years an equal amount of 3D beams requires a longevity of $\sim 100$ Myr to contribute equally to $\bar{k}$. 

As a last consideration, we note that the total birthrate $\Gamma$ in Eqs.~\eqref{NGtot1} and \eqref{ktot1} can be eliminated using the
Drake number $N_D=\Gamma\bar{L}$, yielding: 
\begin{align}
\label{NGtot2}
N_G&=N_D\left(1+\dfrac{t_G}{\bar{L}}\sum_i x_i u_i\right), \\
\label{ktot2}
\bar{k}&=N_D\sum_i  x_iv_i\dfrac{\bar{L}_i}{\bar{L}},
\end{align}
allowing us to translate in terms of $N_G$ and $\bar{k}$ the rich literature devoted to the Drake equation.
By further eliminating $N_D$ from \eqref{NGtot2} and \eqref{ktot2} we get
\begin{equation}
\label{ktot3}
\frac{\bar{k}}{N_G}=\dfrac{\sum_i  x_iv_i\bar{L}_i}{\bar{L}+t_G\sum_i x_i u_i},
\end{equation}
which expresses the fraction of galactic signals crossing Earth in terms of the remaining unknown temporal variables: the longevities.
We note that 
Eq.~\ref{ktot3} generalizes a similar formula derived for the case of isotropic signals in \citet{Grimaldi2018a} and \citet{Grimaldi2018}. 
The two formulas are however not fully equivalent because in those works
 $\bar{k}$ was put in relation to the number of emission processes released during the last $t_G$ years rather
than using the number $N_G$ of emissions physically intersecting the Galaxy.

\section{Conclusions}
\label{concl}

In this paper, we have introduced other statistical quantities than the Drake number $N_D$ to characterize the population
of EM technosignatures in the Milky Way. We have considered the average number of EM emissions present in the Galaxy, 
$N_G$, and the average number $\bar{k}$ of emissions intersecting the Earth (or any other site in the Galaxy). 
Unlike $N_D$, $\bar{k}$ and $N_G$ provide 
measures of the number of emission processes that are not necessarily released by currently active emitters, but that can be
potentially detected on Earth ($\bar{k}$) or that still occupy physically the Galaxy ($N_G$). In order to study how these 
indicators are affected by the signal directionality we have considered  in addition to the case of isotropic emission processes also strongly 
anisotropic ones like narrow beams pointing in random directions and rotating beacons.

Under the assumption that the emission 
birthrates did not change during the recent history of the Galaxy, we have shown that $\bar{k}=N_D$ only for isotropic processes
and for emissions originating from rotating beacons sweeping the galactic disk. In all the other cases considered (beamed signals
directed randomly and lighthouses with tilted rotation axis) $\bar{k}$ can be orders of magnitudes smaller than the Drake number,
showing that $N_D$ may largely overestimate the possible occurrence of signals that can be remotely detected.

We have further discussed at length $N_G$ as the proper indicator of the galactic abundance of technosignatures.
We have shown that $N_G$, leaving aside the special case of narrow beams directed uniformly in 3D space, is only marginally affected 
by the signal directionality. A central result of the present study is that $N_G$ becomes independent of the signal longevity if 
this is shorter than about $10^5$ ly, yielding therefore a measure of the abundance of galactic technosignatures that depends only 
on the emission birthrate. 

\section*{Acknowledgements}
The author thanks Amedeo Balbi and Geoffrey W. Marcy for fruitful discussions.

\section*{Data availability}
The data underlying this article are available in the article.

\appendix

\section{anisotropic emissions}

\subsubsection{$\bar{k}$ for random beams}
\label{kbeams}

Let us consider an emitter located at $\mathbf{r}$ transmitting since a time $t$ before present
a conical beam of aperture $\alpha_0$ \citep{Forgan2014,Grimaldi2017}. During the entire lifetime $L$ of the emission, the beam axis is held oriented 
along the direction of a unit vector $\hat{n}$. The region of space filled by the radiation is the intersection between a cone of apex at $\mathbf{r}$ 
and a spherical shell centred on the cone apex with outer radius $ct$ and thickness $cL$. As done for the isotropic case, we neglect the internal 
structure of this region arising in the case of an intermittent beam. 

\begin{figure}
	\begin{center}
		\includegraphics[width=\columnwidth,clip=true]{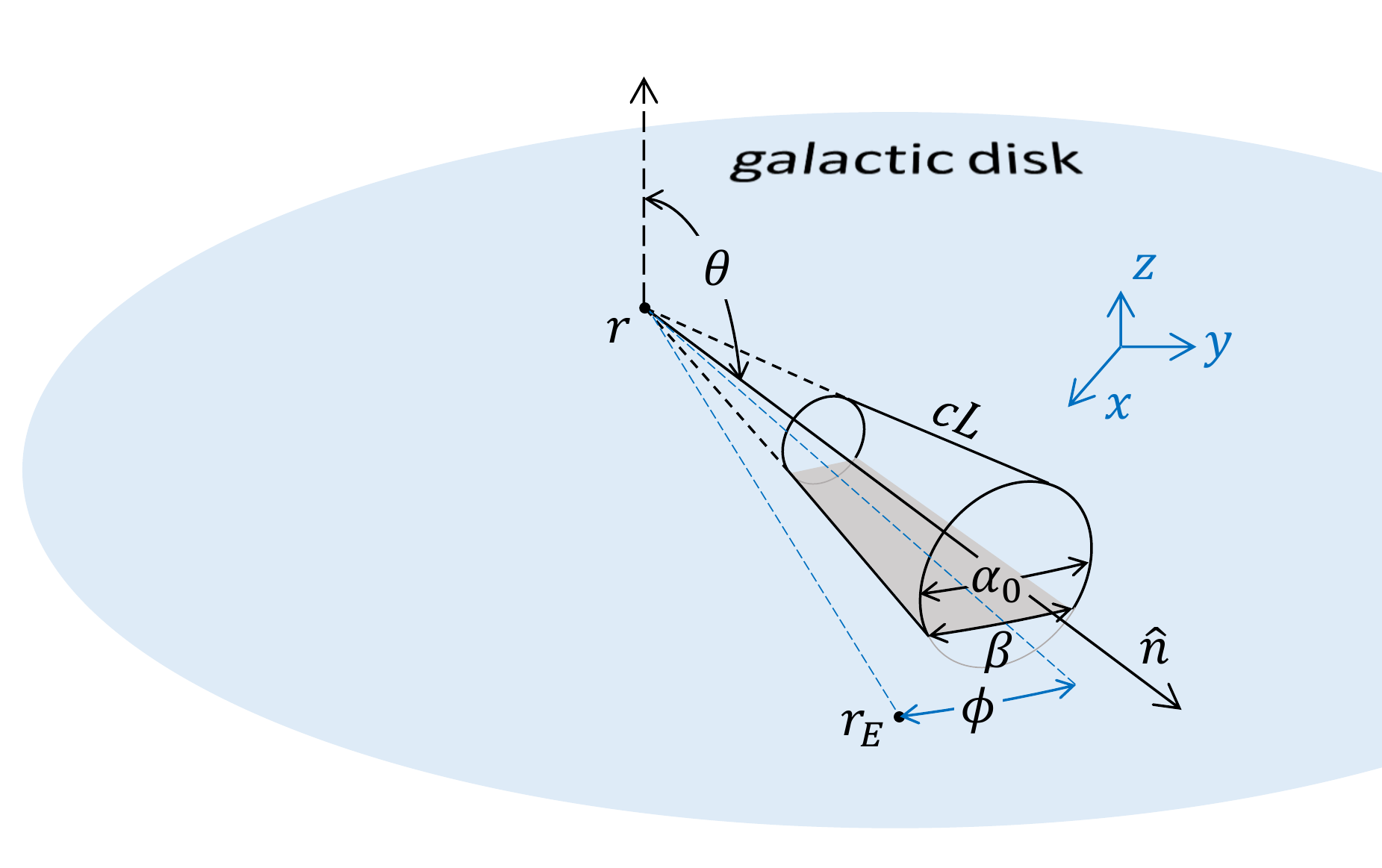}
		\caption{Schematic illustration of a beamed emission of duration $L$ and aperture $\alpha_0$. 
			The Earth and the emitter positions are denoted respectively by the vectors $\mathbf{r}$ and 
			$\mathbf{r}_\textrm{E}$. The beam axis of the lighthouse is oriented along the direction $\hat{n}$ and forms an angle $\theta$ with 
			the $z$-axis. $\phi$ is the angle formed by the direction of $\mathbf{r}-\mathbf{r}_\textrm{E}$ and the projection
			of $\hat{n}$ over the galactic disk.The grey region denotes the overlap area between the galactic disk and the beam.}\label{figA1}
	\end{center}
\end{figure}

The angular sector formed by the overlap of the conical beam with the galactic plane (grey region in Fig.~\ref{figA1}) subtends the angle 
\begin{equation}
	\label{beta}
	\beta=\left\{
	\begin{array}{ll}
		2\arccos\!\left[\dfrac{\cos(\alpha_0/2)}{\sin(\theta)}\right], & \vert\theta-\pi/2\vert\leq \alpha_0/2 \\ [8pt]
		0, & \textrm{otherwise}
	\end{array}\right.
\end{equation}
where $\theta\in[0, \pi]$ is the angle formed by $\hat{n}$ with the $z$-axis. 
From this construction, we se that the beam will cross the Earth if $\mathbf{r}_\textrm{E}$ is located within the angular sector, that is, if Eq.~\eqref{indi1} is satisfied 
and $\vert\phi\vert\leq \beta/2$, where $\phi$ is the angle formed
by $\mathbf{r}_\textrm{E}-\mathbf{r}$ and the projection of $\hat{n}$ on the $x$-$y$ plane, Fig.~\ref{figA1}. 

By adopting a constant birthrate of beamed signals with random orientations of $\hat{n}$, the integration over $t$ under the condition \eqref{indi1} yields 
the factor $L$, as in Eq.~\eqref{k2}. Introducing the random beam (rb) emission rate $\Gamma_\textrm{rb}$
and the corresponding average longevity $\bar{L}_\textrm{rb}$, the mean number of beamed signals crossing Earth reduces therefore to:
\begin{equation}
	\label{k3a}
	\bar{k}_\textrm{rb}(\alpha_0)=\Gamma_\textrm{rb}\bar{L}_\textrm{rb}\int\!d\hat{n}\,g(\hat{n})\Theta(\beta/2-\vert\phi\vert),
\end{equation}
where $d\hat{n}=d\phi d\theta\sin\theta$, $g(\hat{n})$ is the PDF of the direction of $\hat{n}$, and $\Theta(x)=1$ for $x\geq 0$ and $\Theta(x)=0$ or $x<0$ is the
unit step function.

In the case in which the beams are oriented uniformly in three dimensions (3D), the PDF of $\hat{n}$ is $g(\hat{n})=1/4\pi$ and using Eq.~\eqref{beta} the 
integration over $\hat{n}$ yields $\frac{1}{2}[1-\cos(\alpha_0/2)]$, which is simply the fractional solid angle subtended by the beam \citep{Grimaldi2017}.
For beam directions distributed uniformly over the two-dimensional (2D) galactic plane, $g(\hat{n})$ is a Dirac-delta function peaked at $\theta=\pi/2$, 
$g(\hat{n})=\delta(\theta-\pi/2)/2\pi$, and the orientational average reduces simply to $\alpha_0/(2\pi)$. For random 3D and 2D beam orientations
we obtain therefore:
\begin{equation}
	\label{rb0}
	\bar{k}_\textrm{rb}(\alpha_0)=\left\{
	\begin{array}{ll}
		\dfrac{1-\cos(\alpha_0/2)}{2}\Gamma_\textrm{rb}\bar{L}_\textrm{rb}, &  \textrm{3D random beams}, \\[8pt]
		\dfrac{\alpha_0}{ 2\pi}\Gamma_\textrm{rb}\bar{L}_\textrm{rb}, &  \textrm{2D randombeams}.
	\end{array}
	\right.
\end{equation}
Under the assumption that the beams have angular apertures distributed over small values of $\alpha_0$, Eq.~\eqref{rb0} reduces to:
\begin{equation}
	\label{rb1}
	\bar{k}_\textrm{rb}=\langle\bar{k}_\textrm{rb}(\alpha_0)\rangle\simeq\left\{
	\begin{array}{ll}
		\dfrac{\langle\alpha_0^2\rangle}{16}\Gamma_\textrm{rb}\bar{L}_\textrm{rb}, &  \textrm{3D random beams}, \\[8pt]
		\dfrac{\langle\alpha_0\rangle}{ 2\pi}\Gamma_\textrm{rb}\bar{L}_\textrm{rb}, &  \textrm{2D random beams},
	\end{array}
	\right.
\end{equation}
where $\langle\cdots\rangle$ denotes an average over the $\alpha_0$ values.

\subsubsection{$\bar{k}$ for lighthouses}
\label{klighthouses}

\begin{figure}
	\begin{center}
		\includegraphics[width=\columnwidth,clip=true]{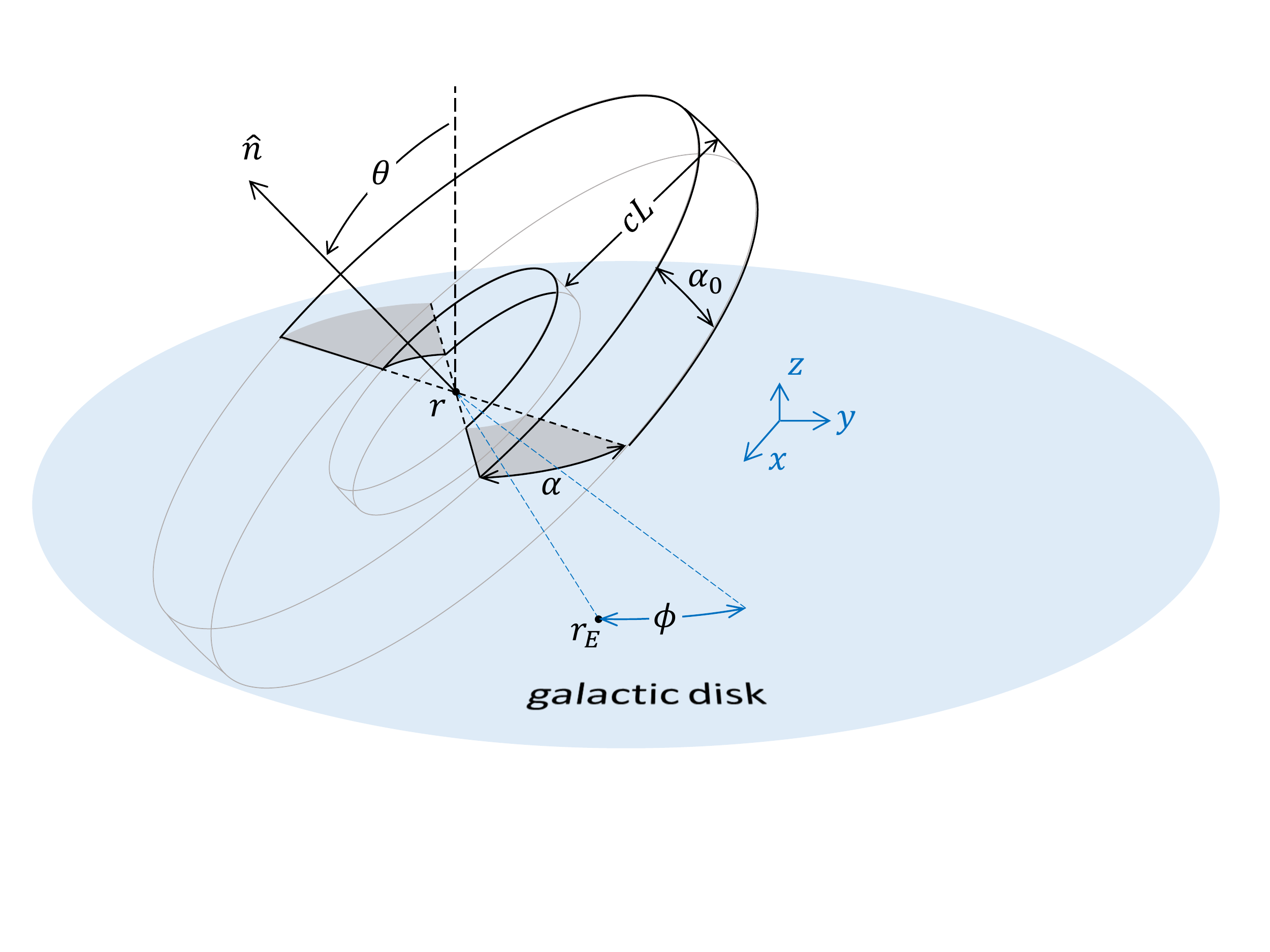}
		\caption{Graphical representation of the effective volume encompassing the regions swept during a time duration $L$ by a lighthouse (or rotating beacon)
			of beam aperture $\alpha_0$. The Earth and the emitter positions are denoted respectively by the vectors $\mathbf{r}$ and 
			$\mathbf{r}_\textrm{E}$. The spin axis of the lighthouse is oriented along the unit vector $\hat{n}$ and forms an angle $\theta$ with 
			the $z$-axis. The grey region denotes the overlap area between the galactic disk and the effective volume.}\label{figA2}
	\end{center}
\end{figure}

We take a lighthouse (lh) located at $\mathbf{r}$ that started transmitting at a time $t$ in the past and for a duration $L$ a conical beam
of angular aperture $\alpha_0$. The effective volume encompassing the region physically filled by the EM radiation
is formed by the overlap between the volume swept by a cone (of aperture $\alpha_0$) rotating about the spin axis $\hat{n}$ with apex at $\mathbf{r}$ and a spherical 
shell concentric to $\mathbf{r}$ of outer radius $ct$ and thickness $cL$ (see Fig.~\ref{figA2}).
The overlap of the effective volume with the galactic plane forms two annular sectors (shown by grey areas in Fig.~\ref{figA2}), symmetric with respect to the line of 
nodes formed by the intersection of the rotation plane with the galactic disk, and subtending the angle
$\alpha$ given by
\begin{equation}
	\label{alfa}
	\alpha=\left\{
	\begin{array}{ll}
		\pi, & 0\leq  \theta\leq \alpha_0/2, \\ [8pt]
		2\arcsin\!\left[\dfrac{\sin(\alpha_0/2)}{\sin(\theta)}\right], & \alpha_0/2<\theta\leq \pi/2,
	\end{array}\right.
\end{equation}
where now $\theta\in[0, \pi/2]$ is the angle formed by $\hat{n}$ with the $z$ direction. As done for the case of random beamed signals, 
the intersection of the angular sector with the Earth is given by the condition \eqref{indi1} and $\vert\phi\vert\leq \alpha/2$, 
where $\phi$ is the angle formed by $\mathbf{r}_\textrm{E}-\mathbf{r}$ and 
the line of nodes (see Fig.~\ref{figA2}). For a constant birthrate of the lighthouses, the average number of randomly distributed spiralling beams 
crossing Earth is thus given by:
\begin{equation}
	\label{sp1}
	\bar{k}_\textrm{lh}(\alpha_0)=\Gamma_\textrm{lh}\bar{L}_\textrm{lh}\int\!d\hat{n}\,g(\hat{n})\Theta(\alpha/2-\vert\phi\vert),
\end{equation}
If $\hat{n}$ is distributed uniformly over the unit sphere (3D case), using Eq.~\eqref{alfa} and $g(\hat{n})=1/4\pi$, the integral over 
$\theta$ reduces exactly to $\sin(\alpha_0/2)$ for $\alpha_0\leq \pi$ and $1$ otherwise. In the case the spin axis is perpendicular 
to the galactic plane (2D case), $g(\theta)$ is a Dirac-delta peak at $\theta=0$, so that the angular average in Eq.~\eqref{sp1} yields $1$. 
Expanding Eq.~\eqref{sp1} for small beam apertures, $\bar{k}_\textrm{lh}=\langle\bar{k}_\textrm{lh}(\alpha_0)\rangle$
reads: 
\begin{equation}
	\label{sp2}
	\bar{k}_\textrm{lh}=\left\{
	\begin{array}{ll}
		\dfrac{\displaystyle\langle\alpha_0\rangle}{\displaystyle 2}\Gamma_\textrm{lh}\bar{L}_\textrm{lh}, & \textrm{3D lighthouses}, \\
		\Gamma_\textrm{lh}\bar{L}_\textrm{lh}, &  \textrm{2D lighthouses}.
	\end{array}\right.
\end{equation}

\begin{figure}
	\begin{center}
		\includegraphics[scale=0.7,clip=true]{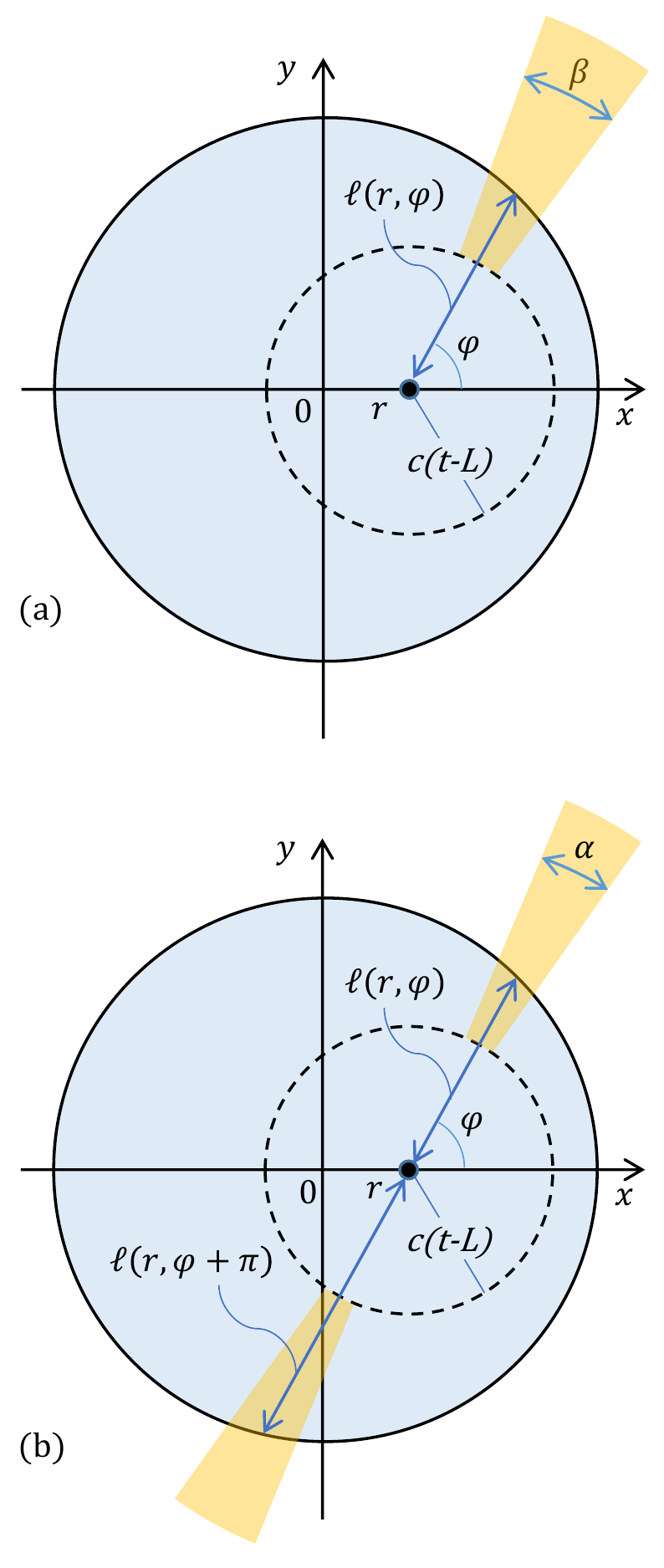}
		\caption{Top view of the galactic disk of radius $R_G$ with superimposed areas covered by emissions of different geometries originated by a source
			located at $(r,0)$. (a): The intersection of a beam with the $x$-$y$ plane forms an annular sector (orange region) subtending
			an angle $\beta$, Eq.~\eqref{beta}, and an inner arc at distance $c(t-L)$ from the emitter. $\ell(r,\varphi)$ is the distance from the emitter to the edge of the Galaxy
			along the direction of the beam axis projected on the $x$-$y$ plane. A narrow beam intersects the galactic disk when $c(t-L)\leq \ell(r,\varphi)$. 
			(b): The two annular sectors of angle $\alpha$, 
			Eq.~\eqref{alfa}, denote the overlap area between the $x$-$y$ plane and the effective volume spanned by a rotating lighthouse with spin axis tilted with respect 
			to the $z$-axis (see Fig.~\ref{figA2}). The intersection of the galactic disk with the rotation plane forms two segments of length $\ell(r,\varphi)$ and 
			$\ell(r,\varphi+\pi)$. The longest of them must be larger than $c(t-L)$ in order for the effective volume to intersect the galactic disk.}\label{figA3}
	\end{center}
\end{figure}

\subsubsection{$N_G$ for random beams}
\label{NGbeams}

In deriving the average number of random beams intersecting the Galaxy, $N_G^\textrm{rb}$, we shall retain only the contributions at the lowest 
order in the beam aperture $\alpha_0$, which simplifies considerably the calculation. 
We take an emitter to be located along the $x$-axis, $\mathbf{r}=(r,0)$, with the 
beam axis directed along $\hat{n}$ forming an azimuthal angle $\varphi$ with $\mathbf{r}$, Fig.~\ref{figA3}(a).
The projection of the beam axis on the $x$-$y$ plane defines the distance 
\begin{equation}
	\label{elle}
	\ell(r,\varphi)=\sqrt{R_G^2-r^2\sin(\varphi)^2}-r\cos(\varphi)
\end{equation}
measured from the emitter position to the edge of the Galaxy (i.e. the circle of radius $R_G$).
As seen from Eq.~\eqref{beta}, the intersection of the beam with the $x$-$y$ plane is non-null only if the polar angle of 
$\hat{n}$ is such that $\vert\theta-\pi/2\vert\leq \alpha_0/2$. Since $\alpha_0\ll 1$, $\hat{n}$ lies approximately on the $x$-$y$ plane and 
a beam emitted at time $t$ for a duration $L$ will intersect the galactic disk only when $c(t-L)$ is smaller than $\ell(r,\varphi)$, as shown in Fig.~\ref{figA3}(a).
Conversely, if $\vert\theta-\pi/2\vert > \alpha_0/2$ the only beams that
intersect the galactic disk are those that are still being transmitted at the present time, that is, those such that $t<L$. After integration over $t$,
$N_G^\textrm{rb}$ at the lowest order in $\alpha_0$ is therefore given by:
\begin{align}
	\label{NGrb1}
	N_G^\textrm{rb}&=\int\!dL\rho_\textrm{rb}(L)\int\! d\mathbf{r}\gamma_\textrm{rb}(\mathbf{r})\int\!d\hat{n}\, g(\hat{n})\nonumber \\
	&\times\left[L+\Theta\left(\frac{\alpha_0}{2}-\left\vert\theta-\frac{\pi}{2}\right\vert\right)\frac{\ell(r,\varphi)}{c}\right]\nonumber \\
	&=\Gamma_\textrm{rb}\bar{L}_\textrm{rb}+t_G\frac{\eta}{\pi}\int\!d\mathbf{r}\gamma_\textrm{rb}(\mathbf{r})E(r/R_G),
\end{align}
where $\eta=\alpha_0/2$ or $\eta=1$ if the direction of $\hat{n}$ is distributed uniformly in 3D space or in the $x$-$y$ plane, and 
$E(x)=\int_0^{\pi/2}\!d\varphi\,\sqrt{1-x^2\sin(\varphi)^2}$ is the complete elliptic integral of the second kind. Using a  
birthrate that is constant over the galactic disk, the integration over $\mathbf{r}$ yields:
\begin{equation}
	\label{NGrb2}
	N_G^\textrm{rb}=\left\{
	\begin{array}{ll}
		\Gamma_\textrm{rb}\left(\bar{L}_\textrm{rb}+\dfrac{\langle\alpha_0\rangle}{2}\dfrac{4}{3\pi}t_G\right), & \textrm{3D random beams} \\[8pt]
		\Gamma_\textrm{rb}\left(\bar{L}_\textrm{rb}+\dfrac{4}{3\pi}t_G\right), & \textrm{2D random beams},
	\end{array}\right.
\end{equation}
from which we see that $N_G^\textrm{rb}$ of narrow beams depends on the angular aperture only in the 3D case.

\subsubsection{$N_G$ for lighthouses}
\label{NGlighthouses}

In the case the spin axis of a rotating beacon is parallel to the $z$-axis, the effective volume
encompassing the radiation (Fig.~\ref{figA2}) intersects the Galaxy as long as $c(t-L)$ is smaller than the maximum distance of the 
emitter from the galactic edge, in full equivalence with the isotropic case. Assuming a spatially uniform birthrate, 
the number of rotating beacon signals intersecting the Milky Way is thus 
$N_G^{lh}=\Gamma_\textrm{lh}(\bar{L}_\textrm{lh}+\frac{5}{6}t_G)$, as in Eq.~\eqref{NGiso2}.

In the more general case in which $\hat{n}$ forms an angle $\theta$ with the $z$-direction, at the lowest order in $\alpha_0$
it suffices to calculate $N_G^{lh}$ by considering the intersection of the rotation plane with the galactic disk, which forms 
an angle $\varphi$ with the $x$-axis. As shown in Fig.~\ref{figA3}(b), the emitter, located at $(r,0)$, cuts the intersection 
line in two segments, generally of different lengths. The longest of these segments has length $\ell_\textrm{max}(r,\varphi)=\ell(r,\varphi+\pi)$
for $0\leq \varphi\leq \pi/2$ and $\ell_\textrm{max}(r,\varphi)=\ell(r,\varphi)$ for $\pi/2\leq \varphi\leq \pi$,  where $\ell(r,\varphi)$ is given in Eq.~\eqref{elle}. 
Since a non-null intersection with the galactic disk is obtained by requiring the inner edge of the effective volume, $c(t-L)$, to be smaller than
$\ell_\textrm{max}(r,\varphi)$, we obtain:
\begin{align}
	\label{NGsp1}
	N_G^\textrm{lh}&=\!\int\!\!dL\rho_\textrm{lh}(L)\!\int\!\! d\mathbf{r}\gamma_\textrm{lh}(\mathbf{r})\!\int\!d\hat{n}\, g(\hat{n})
	\!\left[L+\frac{\ell_\textrm{max}(r,\varphi)}{c}\right]\nonumber \\
	&=\Gamma_\textrm{lh}\bar{L}_\textrm{lh}+\frac{t_G}{\pi}\int\!d\mathbf{r}\gamma_\textrm{lh}(\mathbf{r})[E(r/R_G)+r/R_G],
\end{align}
where we have assumed that the orientation of $\hat{n}$ is distributed uniformly over 3D.
For a spatially uniform $\gamma_\textrm{lh}(\mathbf{r})$ the integration over $\mathbf{r}$ yields $2\Gamma_\textrm{lh}$,
so that for the two cases examined ($\hat{n}$ random and $\hat{n}\parallel z$) $N_G^\textrm{lh}$ reduces to:
\begin{equation}
	\label{NGsp2}
	N_G^\textrm{lh}=\left\{
	\begin{array}{ll}
		\Gamma_\textrm{lh}\left(\bar{L}_\textrm{lh}+\dfrac{2}{\pi}t_G\right), & \textrm{3D lighthouses} \\[8pt]
		\Gamma_\textrm{lh}\left(\bar{L}_\textrm{lh}+\dfrac{5}{6}t_G\right), & \textrm{2D lighthouses}.
	\end{array}\right.
\end{equation}

\bsp	
\label{lastpage}
\end{document}